\documentclass[epj]{svjour}
\usepackage{amssymb}
\usepackage{graphics}
\usepackage{cite}
\usepackage[T1]{fontenc}
\begin{document}
\title{Realizing a stable magnetic double-well potential on an atom chip}

\author{J. Est\`eve, T. Schumm, J.-B. Trebbia, I. Bouchoule,  A. Aspect and C. I. Westbrook}
\institute{Laboratoire Charles Fabry de l'Institut d'Optique, UMR
8501 du CNRS, 91403 Orsay Cedex, France}
\date{Received: date / Revised version: date}
%
\abstract{ We discuss design considerations and the realization of
a magnetic double-well potential on an atom chip using
current-carrying wires. Stability requirements for the trapping
potential lead to a typical size of order microns for such a
device. We also present experiments using the device to manipulate
cold, trapped atoms.
} 
\authorrunning{J. Est\`eve et al.}
\titlerunning{Realizing a stable magnetic double-well potential on an atom chip}

\maketitle

\section{Introduction}

Progress in the fabrication and use of atom chips has been rapid
in the past few years~\cite{Folmanrevue}. Two notable recent
results concern the coherent manipulation of atomic ensembles on
the chip: Ref.~\cite{treutlein:2004} reported the coherent
superposition of different internal degrees of freedom while
in~\cite{wang:2004,zimmermann:private} a coherent beam splitter
and interferometer using Bragg scattering was reported. In the
same vein, the observation of a coherent ensemble in a chip-based
double-well potential also represents a significant milestone. The
dynamics of a Bose-Einstein condensate in a double-well potential
has attracted an enormous amount of theoretical
attention~\cite{theory}, in part because one can thus realize the
analog of a Josephson junction. Indeed, coherent oscillations of
atoms in a laser induced double-well potential have recently been
observed~\cite{albiez:2005}. In addition, the observation of an
oscillation in a double-well amounts to the realization of a
coherent beam splitter which promises to be enormously useful in
future atom interferometers based on atom
chips~\cite{cassetari:2000,hinds:2001,hansel:2001c,andersson:2002}.

In this paper, we discuss progress towards the realization of
coherent oscillations on an atom chip. We begin with some
theoretical considerations concerning atoms in double-well
potentials and show that a configuration with two elongated
Bose-Einstein condensates that are coupled along their entire
length allows one to achieve a variety of oscillation regimes.
Then, we will discuss design considerations which take into
account stability requirements for the trapping potentials in the
transverse direction. Fluctuations in the external magnetic fields
impose a typical size less than or on the order of microns on the
double-well. Atom chips implemented with current-carrying wires
are well suited to elongated geometry and the micron size scale.
After these general considerations, we discuss a particular
realization of a magnetic double-well potential which has been
constructed in our laboratory. Our device has much in common with
the proposal of Ref.~\cite{hinds:2001}, but we believe it
represents an improvement over the first proposal in that it is
quite robust against technical noise in the various currents.  We
will also show some initial observations with the device using
trapped $^{87}$Rb atoms.

\section{Dynamics of two elongated Bose-Einstein condensates coupled by tunneling}
\label{sec.theory}

The dynamics of a Bose-Einstein condensate in a double-well
potential has been widely discussed in the
literature~\cite{theory}. In this section, we will review some of
the basic results and apply them to the case of two elongated
condensates coupled by tunneling along their entire length $L$.

We assume the trapping potential can be written as the sum of a
weakly confining longitudinal potential $V_l(z)$ and a two
dimensional (2D) double-well in the transverse direction
$V_r(\mathbf{r})$. We characterize the two transverse potential
wells by the harmonic oscillator frequency at their centers,
$\omega_0$. We also assume that the longitudinal motion of the
atoms is decoupled from the transverse motion, so that we may
restrict ourselves to a 2D problem. As discussed
in~\cite{bouchoule:2005}, this assumption is not always valid, but
it gives a useful insight into the relevant parameters of the
problem and how they affect the design of the experiment.

Considering the motion of a single atom, the lowest two energy
states of the 2D potential $V_r$ are symmetric and antisymmetric
states, $|\phi_s\rangle$ and $|\phi_a\rangle$. The energy
splitting $\hbar \, \delta$ between them is related to the
tunneling matrix element between the states describing a particle
in the right and left wells, $|\phi_r\rangle$ and
$|\phi_l\rangle$. When including atom-atom interactions, we assume
that the longitudinal linear density $n_1$ is low enough to
satisfy $n_1 \, a \ll 1$ where $a$ is the $s$-wave scattering
length. In this case, the interaction energy is small compared to
the characteristic energy $\hbar \, \omega_0$ of the transverse
motion, and in a mean field approximation the eigenstates of the
Gross-Pitaevskii equation are identical to the single particle
eigenstates. The tunneling rate $\delta$ is unchanged in this
approximation. If in addition, $\delta \ll \omega_0$, the two mode
approximation in which one considers only the states
$|\phi_s\rangle$ and $|\phi_a\rangle$ (or $|\phi_l\rangle$ and
$|\phi_r\rangle$) is valid.

We now define two characteristic energies $E_J$ and $E_C$. The
Josephson energy $E_J=N \, \hbar \, \delta /2$ characterizes the
strength of the tunneling between the wells. The charging energy
$E_C=4\, \hbar \, \omega_0 \, a/L$ is analogous to the charging
energy in a superconducting Josephson junction and characterizes
the strength of the inter atomic interaction in each well. The
properties of the system depend drastically on the ratio
$E_C/E_J$~\cite{theory}. For the considered elongated geometry,
this ratio is equal to $(4 \, n_1 \,a) \times (\omega_0/\delta)
\times 1/N^2$. A ratio $\omega_0 / \delta$ of 10 is enough to
insure the validity of the two mode approximation. We have also
assumed $n_1 \, a \ll 1$ and since $N \gg 1$ we indeed obtain
$E_C/E_J \ll 1$. This means that the phase difference between the
two wells is well defined and that a mean field description of the
system is valid.

In the mean field approximation, the transverse part of the atomic
wavefunction can be written $|\phi\rangle=c_l \, |\phi_l\rangle +
c_r \, |\phi_r\rangle$ where $c_l$ and $c_r$ are complex numbers.
The atom number difference $\Delta N=(|c_l|^2-|c_r|^2)/2$ and the
phase difference $\Delta \theta = \arg(c_l/c_r)$ evolve as the
classically conjugate variables of a non rigid pendulum
Hamiltonian~\cite{smerzi:1997}. The solutions of the motion for
$\Delta N$ and $\Delta \theta$ have been analytically
solved~\cite{smerzi:1997,raghavan:1999}. Depending on the ratio
$E_J/(N^2 \, E_C)$, we distinguish two regimes: the Rabi regime
($E_J \gg N^2 \, E_C$) and the Josephson regime ($E_J \ll N^2 \,
E_C$). For a fixed geometry ($L$, $\delta$ and $\omega_0$ fixed),
the Rabi regime is delimited by $N \ll N_C$ where $N_C=\delta \,
L/(4 \, \omega_0 \,a)$ while the Josephson regime corresponds to
$N \gg N_C$. For a box like potential of length $L=1$~mm and a
ratio $\omega_0/\delta=10$, this number corresponds to $N_C=5000$
for $^{87}$Rb atoms.

In the Rabi regime, an initial phase difference of $\pi/2$ leads
to the maximal relative atom number difference $\Delta N/N = 1/2$.
In the Josephson regime, the signal $\Delta N/N$ is limited to
$\sqrt{N_C/N}$. One motivation to attain the Rabi regime is to
maximize the relative population difference. If tunneling is to be
used as a beam splitting device in an atom interferometer, the
Rabi regime is clearly favorable as it maximizes the measured
signal. It is also important to note that the neglect of any
longitudinal variations in the atom number difference or the
relative phase is only valid deep in the Rabi
regime~\cite{bouchoule:2005}.


The specific geometry of two elongated Bose-Einstein condensates
coupled by tunneling is of special interest since it allows one to
tune the strength of the interaction compared to the tunneling
energy by adjusting the longitudinal atomic density. This allows
realization of experiments in both the Rabi regime and the
Josephson regime. On the other hand a complication of the
elongated geometry is the coupling between the transverse and
longitudinal motions introduced by interactions between atoms.
This coupling is responsible for dynamical longitudinal
instabilities in presence of uniform Josephson
oscillations\cite{bouchoule:2005}. However,
Ref.~\cite{bouchoule:2005} predicts that a few Josephson
oscillations periods should be visible before instabilities become
too strong. Furthermore, the study of these instabilities may
prove quite interesting in their own right. Other manifestations
of the coupling between the transverse and the longitudinal motion
may be observed. In particular, Josephson vortices are expected
for large linear density\cite{Kaurov:2005}. These nonlinear
phenomena are analogous to observations on long Josephson
junctions in superconductors~\cite{likharev:book}.

\section{Realization of a magnetic double-well potential}
\begin{figure}
\centering
  \includegraphics{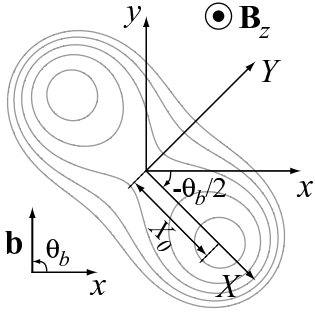}
\caption{Adding a small magnetic field $\mathbf{b}$ to the
magnetic hexapole described in Eq.~\ref{eq.hexapole} will split
the hexapole into two quadrupoles. The distance between the two
minima increases with $b$ and the direction along which the minima
are split depends on the orientation of $\mathbf{b}$ as shown in
the figure. We also have plotted lines of constant modulus of the
total magnetic field (equipotential lines for the atoms).}
\label{fig.splitting}       
\end{figure}

We now turn to some practical consideration concerning the
realization of the transverse double-well potential $V_r(x,y)$
using a magnetic field. As first pointed out in~\cite{hinds:2001},
a hexapolar magnetic field is a good starting point to produce
such a potential. The hexapolar field can be written
\begin{equation}\label{eq.hexapole}
    \left\{\begin{array}{cccccc}
      B_x & = & A \, (y^2-x^2) & = & -A \, r^2 \, \cos 2\theta\\
      B_y & = & 2 \, A \, x \, y & = & A \, r^2 \, \sin 2\theta \, .\\
    \end{array}
    \right.
\end{equation}
where $A$ is a constant characterizing the strength of the
hexapole. In the following, we write this constant $A=\alpha \,
\mu_0 \, I /(4 \, \pi \, d^3)$ where $I$ is the current used to
create the hexapole, $d$ is the typical size of the current
distribution creating the magnetic field (see
Fig.~\ref{fig.DispositionFils}) and $\alpha$ is a geometrical
factor close to unity. Adding a uniform transverse magnetic field
$\mathbf{b}=b(\cos \theta_b \, \mathbf{x} + \sin \theta_b \,
\mathbf{y})$ will split the hexapole into two quadrupoles, thus
realizing a double-well potential. The two minima are separated by
a distance $2 \,X_0$ where $X_0=\sqrt{b/A}$ and are located on a
line making an angle $\theta_b/2$ with the $\mathbf{x}$-axis (see
Fig.~\ref{fig.splitting}). Tilting the axis of the double-well
allows one to null the gravitational energy shift which arises
between the two wells if they are not at the same height. This
shift has to be precisely cancelled to allow the observation of
unperturbed phase oscillations in the double-well. For example, if
the two wells are separated by a vertical distance of 1~$\mu$m the
gravitational energy shift between the two wells leads to a phase
difference of 13~rad.ms$^{-1}$ for $^{87}$Rb.

In the rotated basis $(O,X,Y)$ (see Fig.~\ref{fig.splitting}), the
modulus of the total magnetic field is
\begin{equation}
    \mathbf{B}^2 = 2 \, A \, b \, Y^2 +2\, A \, X^2 \, Y^2 +
    A^2(X^2-X_0^2)^2 \, .
\end{equation}
To prevent Majorana losses around each minimum, a uniform
longitudinal magnetic field $B_z$ is added. Under the assumption
$B_z \gg b$, the potential seen by an atom with a magnetic moment
$\mu$ and a mass $m$ is
\begin{equation}\label{eq.potentiel}
    V(X,Y)  \simeq \frac{m \, \omega_0^2}{4} Y^2 + \frac{m \, \omega_0^2}{4\, X_0^2} X^2 \,
    Y^2 + \frac{m \, \omega_0^2}{8\, X_0^2}(X^2-X_0^2)^2
\end{equation}
with $\omega_0=\sqrt{4 \, \mu \, A \, b/(m \, B_z)}$.

\begin{figure}
\centering
\resizebox{0.75\columnwidth}{!}{
  \includegraphics{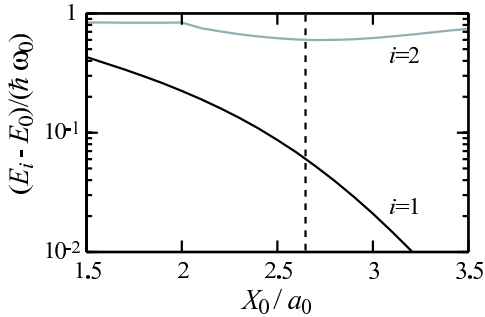}
} \caption{Bohr frequency between the ground state and the first
two excited states of the double-well potential versus the spacing
between the wells. The dashed line corresponds to $X_0=2.65 \,
a_0$ for which the Bohr frequency $\omega_{2,0}$ is ten times
bigger than the tunneling rate $\delta=\omega_{1,0}$.}
\label{fig.energies}       
\end{figure}

Around each minimum $(X=\pm X_0,Y=0)$, the potential is locally
harmonic with a frequency $\omega_0$ and we denote
$a_0=\sqrt{\hbar/(m\,\omega_0)}$ the size of the ground state of
this harmonic oscillator. On the $X$-axis, we recover the 1D
double-well potential usually assumed in the literature. As seen
in equation~(\ref{eq.potentiel}), the potential is entirely
determined by the values of $\omega_0$ and $X_0$. We have computed
the energy differences between the ground state and the two first
excited states for a single atom as a function of these two
parameters (see Fig.~\ref{fig.energies}). The Bohr frequency
$\omega_{1,0}$ is equal to the tunneling rate $\delta$. We
calculate that a ratio $X_0/a_0=2.65$ ensures that $\omega_{2,0} =
10 \, \delta$, so that the two mode approximation is valid.
Further calculations are made for a double-well potential
fulfilling the condition $X_0/a_0=2.65$.

\subsection{Stability of the double-well}
We now turn to the analysis of the stability of the system with
respect to fluctuations of magnetic field. We will impose two
physical constraints: first we require a stability of 10\% on the
tunneling rate $\delta$ and second we impose a gravitational
energy shift between the two wells of less than 10\% of the
tunneling energy. Assuming a perfectly stable hexapole and that
fluctuations in the external fields can be kept below 1~mG, we
will obtain constraints on the possible size of the current
distribution $d$ and on the spacing between the two wells $X_0$.

The geometry of the magnetic double-well is determined by four
experimental parameters: $I$ the current creating the hexapole,
$d$ the size of the current distribution, $B_z$ the longitudinal
magnetic field and $b$ the transverse field. To minimize the
sensitivity of the system to magnetic field fluctuations, the
current $I$ creating the hexapole should be maximized. If we
suppose the wires that create the hexapole are part of an atom
chip, the maximal current allowed in such wires before damage
scales as $I=I_0 (d/d_0)^{3/2}$~\cite{groth:2004}. Henceforth we
suppose that the current $I$ follows this scaling law and is not a
free parameter anymore. Furthermore, the condition of the last
section $X_0/a_0=2.65$ relates $b$ and $B_z$. Thus we are left
with only two free parameters which may be chosen as the size of
the source $d$ and the distance between the wells $X_0$. The
experimental parameters $b$ and $B_z$ can be deduced afterwards.

\begin{figure}
\centering
\resizebox{0.75\columnwidth}{!}{
  \includegraphics{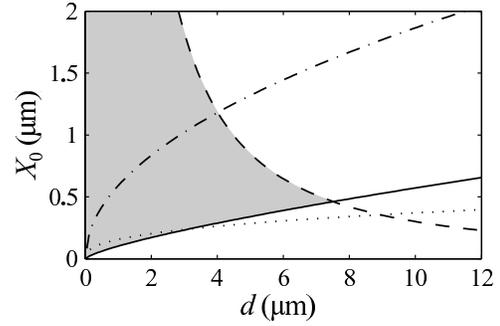}
} \caption{ Stability diagram for the size $d$, and the well
separation $X_0$. The gray area represents pairs of $d$ and $X_0$
for which a 1~mG fluctuation of an external magnetic field does
not significantly disturb the double-well. The solid line delimits
the zone where the tunneling rate fluctuates less than 10\%, while
the dashed line shows the area below which the fluctuations of the
gravitational energy are less than 10\% of the tunneling rate. In
addition we show two additional constraints because of $B_z$: the
area above the dotted line assures a $B_z$ large enough to avoid
Majorana losses and the dash-dotted line corresponds to $B_z <
100$~G. The device we describe in Sec.~\ref{sec.5wires} operates
at $d=5$~$\mu$m and $X_0=0.5$~$\mu$m.}
\label{fig.ScalingLaws} 
\end{figure}

We first calculate the variation $\Delta \delta$ of the tunneling
rate due to longitudinal and transverse magnetic field
fluctuations (respectively noted $\Delta B_z$ and $\Delta b$).
From the numerical calculation of the tunneling rate shown in
figure~\ref{fig.energies} we obtain
\begin{eqnarray}
    \frac{\Delta \delta}{\delta} & = & -2.40 \, \frac{\Delta X_0}{X_0} -2.18 \, \frac{\Delta
    \omega_0}{\omega_0} \\
    & = & -4.27 \, \frac{\Delta b}{b} + 1.09 \, \frac{\Delta
    B_z}{B_z} \, . \label{eq.DeltaFluct}
\end{eqnarray}
If we require a relative stability of 10\% on the tunneling rate,
the required relative stability for the magnetic fields $b$ and
$B_z$ is approximately the same and is easily achievable with a
standard experimental setup. However fluctuations due to the
electromagnetic environment may be problematic. Within the
assumption $B_z \gg b$, only the first term in
equation~(\ref{eq.DeltaFluct}) contributes. Thus we have
\begin{equation}
    \frac{\Delta \delta}{\delta} \simeq -4.27 \, \frac{4\, \pi d^3}{\alpha \, \mu_0 \, I \, X_0^2} \Delta
    b \, .
\end{equation}
Using the scaling law stated above for the current creating the
hexapole, we finally obtain the following expression for the
tunneling rate fluctuations due to variations of the transverse
magnetic field
\begin{equation}
    \frac{\Delta \delta}{\delta} \simeq -4.27 \, \left(\frac{4 \, \pi \, d_0^{3/2}}{\alpha \, \mu_0 \, I_0}\right)
    \frac{d^{3/2}}{X_0^2} \, \Delta b \, . \label{eq.condition1}
\end{equation}
We require a relative stability of 10\% for the tunneling rate
given an amplitude $\Delta b=1$~mG for the magnetic field
fluctuations. This limits the possible values for $X_0$ and $d$ to
the domain above the continuous line plotted in
figure~\ref{fig.ScalingLaws}. For the numerical calculation, we
have used a geometrical factor $\alpha=4/\sqrt{3}$ (see
Sect.~\ref{sec.5wires}) and the following values for $I_0$ and
$d_0$. A maximal current of $I_0=20$~mA is reasonable for gold
wires having a square section of 500~nm$\, \times \,$500~nm which
are deposited on an oxidized silicon wafer~\cite{groth:2004}. Such
wires can be used in a configuration where the typical distance
between wires is $d_0=5$~$\mu$m.

We now turn to the calculation of the fluctuations of the
gravitational energy shift between the two wells. Transverse
magnetic field fluctuations lead to fluctuations $\Delta h= \Delta
b/(A \, X_0)$ of the height difference between the two wells. The
associated fluctuations of the gravitational energy difference
have to be small compared to the tunneling energy so that the
phase difference between the wells is not significantly modified
during one oscillation in the double-well. The ratio between these
two energies is
\begin{equation}
    \frac{m\, g \, \Delta h}{\hbar \, \delta} \simeq 2.37 \, \left(\frac{4\, \pi \, m^2 \, g \, d_0^{3/2}}{\alpha \, \mu_0 \, I_0 \, \hbar^2 }\right) X_0 \, d^{3/2} \,\Delta
    b\, .
\end{equation}
We have used the same scaling law as before for the current in the
hexapole. The possible values for $X_0$ and $d$ that insure this
ratio being smaller than 10\% are located below the dashed line in
figure~\ref{fig.ScalingLaws}. We have assumed the same numerical
parameters as for the first condition. The intersection of the two
possible domains we have calculated for $X_0$ and $d$ corresponds
to the gray area in figure~\ref{fig.ScalingLaws}. The main result
is that the characteristic size of the source $d$ has to be
smaller than 7.5~$\mu$m in order to achieve a reasonable stability
of the double-well. This motivates the use of atom chips to create
a magnetic double-well where external magnetic fluctuations of
1~mG still allows the possibility of coherently splitting a
Bose-Einstein condensate using a magnetic double-well potential.

We have also plotted in figure~\ref{fig.ScalingLaws} the limit
(dotted line) above which the condition $\mu \, B_z >10 \, \hbar
\, \omega_0$ is fulfilled. This insures the Majorana loss to be
negligible in the double-well. Furthermore, above this line the
condition $B_z \gg b$ which is assumed in all our calculation is
also fulfilled. We see this condition is not very restrictive and
does not significantly reduce the domain of possible parameters.
However we note that this condition becomes the limiting factor as
one decreases the size $d$ of the current distribution. The last
plotted dash-dotted line delimits the more practical usable
parameters. Above this line the longitudinal field $B_z$ is
greater than 100~G. Such high values of the longitudinal field
should be avoided since the longitudinal field may have a small
transverse component that would disturb the double-well.

\section{Experimental realization of a magnetic double-well on an atom
chip}\label{sec.5wires} 
\begin{figure}
\centering
\resizebox{0.75\columnwidth}{!}{
  \includegraphics{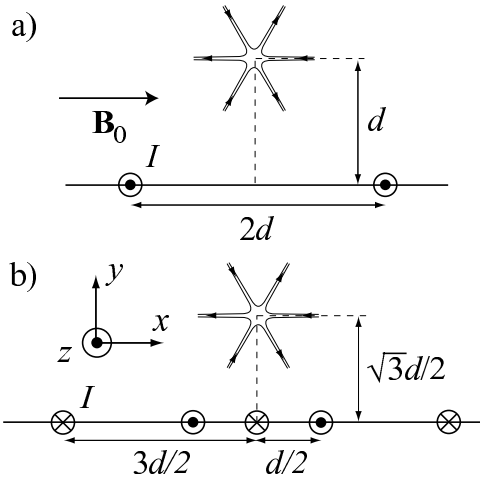}
} \caption{Two configurations that produce a hexapolar magnetic
field. Each wire carries the same current $I$. In (a) the hexapole
is obtained with two wires and a uniform magnetic field $B_0=\mu_0
\, I /(2 \, \pi \, d)$. In (b) it is produced by 5 wires and no
external field. In the first configuration the stability of $B_0$
relative to $I$ is critical. The second configuration avoids this
difficulty provided that the wires are connected in series.}
\label{fig.DispositionFils}       
\end{figure}

As first proposed in~\cite{hinds:2001},
the simplest scheme to obtain a hexapolar magnetic field on an
atom chip uses two wires and an external uniform field (see
Fig.~\ref{fig.DispositionFils}.a). Denoting $2\, d$ the distance
between the two wires, the value of the external field has to be
$B_0=\mu_0 \, I/(2\, \pi \, d)$. One then obtains a hexapole
located at a distance $d$ from the surface of the chip. This
configuration leads to a geometrical factor $\alpha=1$. In order
to safely lie in the stability domain in
figure~\ref{fig.ScalingLaws}, one can choose $d=5$~$\mu$m. This
leads to a current $I=20$~mA and to a uniform magnetic field
$B_0=8$~G. The required relative stability $\Delta B_0/B_0$ for
this field is about $10^{-4}$ since fluctuations of only 1~mG are
tolerable \footnote{More precisely the ratio $I/B_0$ has to be
kept constant with such accuracy. Here we assume that the current
$I$ in the wires does not fluctuate.}. Relative temporal stability
of this magnitude can be achieved with the appropriate
experimental precaution, but it is quite difficult to produce a
spatially homogeneous field on the overall length of the
condensate (1~mm) with such accuracy.

To circumvent this difficulty we propose to realize the hexapolar
field using only wires on the chip. Assuming all the wires are
fabricated on the same layer, at least five wires must be used to
create a hexapole. As seen in figure~\ref{fig.DispositionFils},
the distance between the wires can be chosen so that a hexapole is
obtained with the same current running in all the wires. This
allows rejection of the noise from the power supply delivering the
current $I$. For this geometry, we calculate $\alpha=4/\sqrt{3}$
which is the value we used to plot the curves in
figure~\ref{fig.ScalingLaws}.

\begin{figure}
\centering \resizebox{0.75\columnwidth}{!}{
  \includegraphics{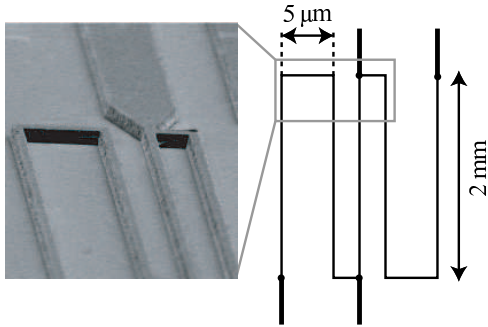}
} \caption{Schematic and SEM picture of our five wire device. The
design allows one to send the same current with a single power
supply in all the wires to create a magnetic hexapole. The
connections on the central wires allow us to imbalance the
currents between the central wire, the two left wires and the two
right wires.}
\label{fig.ChipImage}       
\end{figure}

We have implemented this five wire scheme on an atom chip. The
wires are patterned using electron beam lithography on an oxidized
silicon wafer covered with a 700~nm thick evaporated gold layer.
Each wire has a 700~nm$\, \times \,$700~nm cross-section and is
2~mm long. Figure~\ref{fig.ChipImage} shows the schematic diagram
of the chip and a SEM image of the wire ends. This design allows
us to send the same current in the five wires using a single power
supply. The extra connections are used to add a current in the
central wire in order to split the hexapole into two quadrupoles
without any external magnetic field. We can also change the
current in the left (right) pair of wires in order to release the
atoms from the left (right) trap when the separation between the
wells is large enough. The transverse wires connecting the five
wires at their ends insure the longitudinal confinement of the
atoms in a box like potential.

The distance $d$ characterizing the wire spacing is 5~$\mu$m.
Using the exact expression of the magnetic field created by the
five wires, we have carried out numerical calculation of the
spectrum of the double-well. Using a transverse field $b=60$~mG
and a longitudinal field $B_z=550$~mG, we obtain a spacing between
the wells of $2\, X_0=1.0$~$\mu$m and a tunneling rate of
$\delta=2\pi \times 290$~Hz. The parameters have been chosen to
fulfill the condition $\omega_{0,2} = 10 \, \delta$ and to lie in
the center of the stability domain. We have checked numerically
that the two conditions on the stability of the tunneling rate and
of the gravitational energy shift are indeed fulfilled.

\subsection{Splitting of a thermal cloud}
In order to load the double-well with a sample of cold $^{87}$Rb
atoms, the five wire chip is glued onto an atom chip like that
used in a previous experiment to produce a Bose-Einstein
condensate~\cite{esteve:2004}. The five wire chip surface is
located approximately 150~$\mu$m above the surface of the other
chip. This two-chip design allows one to combine wires having very
different sizes (typically 50~$\mu$m $\times$ 10~$\mu$m for the
first chip and 700~nm $\times$ 700~nm for the five wire chip) and
therefore different current-carrying capacities in a single
device. Large currents are needed to efficiently capture the atoms
from a MOT in the magnetic trap.

Using evaporative cooling, we prepare a sample of cold atoms in a
Ioffe trap created by a Z-shaped wire on the first chip and a
constant external field. Transfer of the atoms to the double-well
potential is achieved by ramping down the current in the Z-shaped
wire and the external field while we ramp up the currents in the
five wires. The final value of the current in the central wire is
smaller (10.4~mA) than for the one in the other wires (17.5~mA).
We use the fact that an imbalanced current in the central wire is
qualitatively equivalent to adding an external transverse field to
the hexapole. Ignoring the field due to the lower chip, these
current values lead to two trapping minima located on the
$y$-axis. The position of the upper minimum is superimposed on the
position of the Ioffe trap due to the lower chip. We typically
transfer of order $10^4$ atoms having a temperature below
1~$\mu$K.

\begin{figure}
\centering \resizebox{0.75\columnwidth}{!}{
  \includegraphics{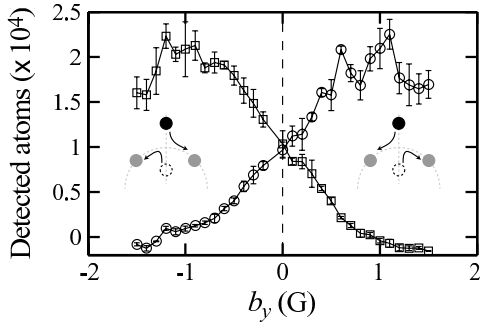}
} \caption{Final number of atoms in the right well ($\square$) and
in the left well ($\circ$) after a splitting experiment. The
schematics depict the trajectories of the two traps during the
sequence. Initially all the atoms are in the upper trap, depending
on the sign of the $y$ component of $b$, the atoms preferentially
end in the left or in the right well. The minimal distance between
the traps depends on the modulus of $b_y$. This distance is zero
if $b_y=0$, leading to a splitting with half of the atoms in each
well.}
\label{fig.ExpSplitting}       
\end{figure}

To realize a splitting experiment, we then increase the current in
the central wire to 17.5~mA and decrease the current in the other
wires to 15~mA. The duration of the ramp is 20~ms. If the external
transverse field is zero, the two traps located on the $y$-axis
coalesce when all the currents are equal and then split along the
$x$-axis when the current in the central wire is above the one in
the other wires. Then, by lowering the current in the left (right)
wires to zero, we eliminate the atoms in the left (right) trap and
measure the number of atoms remaining in the other trap using
absorption imaging. If the external magnetic field has a small
component along the $y$-axis, the coalescence point is avoided and
the atoms initially in the upper trap preferentially go in the
right (left) trap if $b_y$ is positive (negative). The number of
atoms in the left or in the right well as a function of $b_y$ is
plotted in figure~\ref{fig.ExpSplitting}. As expected, we observe
a 50\% split between the two wells if the two traps coalesce using
$b_y=0$. For an amplitude of the magnetic field $b_y$ larger than
0.6~G, the transferred fraction of atoms is almost zero. For this
specific value of the transverse magnetic field, the atomic
temperature at closest approach between the wells is estimated to
be 420~nK. On the other hand, for this transverse field and for
the longitudinal field $B_z=1$~G used in the experiment, the
barrier height between wells at closest approach is 12~$\mu$K.
Thus, the value of the atomic temperature seems too small to
explain our observations. The estimated atomic temperature is
calculated knowing the initial temperature (220~nK) and assuming
adiabatic compression. We have reason to be confident in the
adiabaticity because the temperature is observed to be constant
when the splitting ramp is run backward and forward at
$b_y=0.6$~G.  More precisely, numerical calculations of the
classical trajectories during the splitting indicate that the
typical width of the curves shown on figure~\ref{fig.ExpSplitting}
is approximately three times too large. For the moment, we do not
have a satisfactory explanation for this broadening.

\subsection{Longitudinal potential roughness}
For our present setup, the actual longitudinal potential differs
from the ideal box-like potential because of distortions in the
current distribution inside the
wires~\cite{esteve:2004,Schumm:2005}. Preliminary measurements
indicate a roughness with a rms amplitude of a few mG and a
correlation length of a few $\mu$m. The condensate will thus be
fragmented. Each fragment will be trapped in a potential with a
typical longitudinal frequency of about 400~Hz. Given the same
number of atoms and the same total length for the whole
condensate, the longitudinal density in each fragment will be
approximately ten times higher than for the ideal box-like
potential. Thus the Rabi regime may be out of reach with our
present setup. More precise measurements of the exact longitudinal
potential shape are in progress to determine the maximum ratio
$E_J/(N^2 \, E_C)$ we can actually achieve. Improved wire
fabrication techniques may allow us to obtain a flatter
longitudinal potential and to increase the $E_J/(N^2 \, E_C)$
ratio.

\section{Conclusion}
We have shown that atom chip based setups are well suited to
produce a stable magnetic double-well potential. Our main argument
is that atom chips allow one to design a current distribution
having a characteristic size small enough so that oscillations of
a condensate between the wells can be reproducible despite a noisy
electromagnetic field environment.

We have fabricated a device using five wires spaced by a distance
of a few microns. The preliminary data in
Fig.~\ref{fig.ExpSplitting} show that we have good control over
our transverse magnetic potential, although we cannot entirely
validate our design choices before having observed coherent
oscillations. To do this it remains to reproducibly place a
condensate in the trap so that the two mode description applies
and can be tested.

This work was supported by the E.U. under grants IST-2001-38863
and MRTN-CT-2003-505032 and by the D.G.A. (03.34.033).


%


\begin{thebibliography}{10}

\bibitem{Folmanrevue}
R. Folman, P. Kr\"uger, J. Schmiedmayer, J. Denschlag, and C.
Henkel, Adv. Atom. Mol. Opt. Phys. {\bf 48},  263  (2002), and
references therein.

\bibitem{treutlein:2004}
P. Treutlein, P. Hommelhoff, T. Steinmetz, T.~W. H\"ansch, and J.
Reichel, Phys.~Rev.~Lett. {\bf 92},  203005  (2004).

\bibitem{wang:2004}
Y.-J. Wang, D.~Z. Anderson, V.~M. Bright, E.~A. Cornell, Q. Diot,
T. Kishimoto, M. Prentiss, R.~A. Saravanan, S.~R. Segal, and S.
Wu, cond-mat/0407689 (2004).

\bibitem{zimmermann:private}
C. Zimmermann, Private communication  (2005).

\bibitem{theory}
F. Dalfovo, S. Giorgini, L. Pitaevskii, and S. Stringari, Rev.
Mod. Phys. {\bf 71},  463  (1999), and references therein. A.~J.
Legget, Rev. Mod. Phys. {\bf 73}, 307 (2001), and references
therein.

\bibitem{albiez:2005}
M. Albiez, R. Gati, J. Foelling, S. Hunsmann, M. Cristiani, and
M.~K. Oberthaler, cond-mat/0411757 (2005).

\bibitem{cassetari:2000}
D. Cassettari, B. Hessmo, R. Folmann, T. Maier, and J.
Schmiedmayer, Phys.~Rev.~Lett. {\bf 85},  5483  (2000).

\bibitem{hinds:2001}
E.~A. Hinds, C.~J. Vale, and M.~G. Boshier, Phys.~Rev.~Lett. {\bf
86},  1462 (2001).

\bibitem{hansel:2001c}
W. H\"{a}nsel, J. Reichel, P. Hommelhoff, and T.~W. H\"{a}nsch,
Phys.~Rev.~A {\bf 64},  063607  (2001).

\bibitem{andersson:2002}
E. Andersson, T. Calarco, R. Folman, M. Andersson, B. Hessmo, and
J. Schmiedmayer, Phys.~Rev.~Lett. {\bf 88},  100401  (2002).

\bibitem{bouchoule:2005}
I. Bouchoule, physics/0502050  (2005).

\bibitem{Kaurov:2005}
V. M. Kaurov and A. B. Kuklov, Phys.~Rev.~A {\bf 71}, 011601(R)
(2005).

\bibitem{smerzi:1997}
A. Smerzi, S. Fantoni, S. Giovanazzi, and S.~R. Shenoy,
Phys.~Rev.~Lett. {\bf 79},  4950  (1997).

\bibitem{raghavan:1999}
S. Raghavan, A. Smerzi, and V.~M. Kenkre, Phys.~Rev.~A {\bf 59},
620  (1999).

\bibitem{likharev:book}
K.~K. Likharev, {\em Dynamics of Josephson Junctions and Circuits}
(Gordon and Breach Science Publishers, New York, 1986).

\bibitem{groth:2004}
S. Groth, P. Kr\"uger, S. Wildermuth, R. Folman, T. Fernholz, J.
Schmiedmayer, D. Mahalu, and I. Bar-Joseph, Appl.~Phys.~Lett. {\bf
85},    (2004).

\bibitem{esteve:2004}
J. Est\`eve, C. Aussibal, T. Schumm, C. Figl, D. Mailly, I.
Bouchoule, C.~I. Westbrook, and A. Aspect, Phys.~Rev.~A {\bf 70},
043629  (2004).

\bibitem{Schumm:2005}
T. Schumm, J. Est\`eve, C. Figl, J.-B. Trebbia, C. Aussibal, H.
Nguyen, D. Mailly, I. Bouchoule, C.I. Westbrook and A. Aspect,
Eur.~Phys.~J.~D {\bf 32}, 171 (2005).

\end{thebibliography}

\end{document}